\documentclass[twocolumn,ngerman]{article}
\pdfoutput=1

\usepackage[utf8]{inputenc}
\usepackage[big]{dgruyter}
\usepackage{microtype}
\usepackage{csquotes}
\usepackage[maxcitenames=2,maxbibnames=99,backend=bibtex]{biblatex}
\AtEveryBibitem{\clearlist{language}\clearfield{note}}
\AtEveryBibitem{\clearfield{isbn}}
\usepackage{hologo}
\usepackage{listings}
\begin{document}

\pagestyle{empty}
\twocolumn[
\begin{@twocolumnfalse}
{\Large  Veröffentlicht in \emph{at–Automatisierungstechnik 2023; 71(3): 209–218}}. \\ \\ 
{\Large Die Verlagsveröffentlichung ist verfügbar unter www.degruyter.com\\ DOI: \href{https://doi.org/10.1515/auto-2022-0164}{10.1515/auto-2022-0164}} \\ \\ 
Zu zitieren als:

\vspace{0.1cm}
\noindent\fbox{%
    \parbox{\textwidth}{%
        Graubohm,~R., Loba,~M., Nolte,~M. \& Maurer,~M. (2023). Identifikation auslösender {Umstände} von {SOTIF}-{Gefährdungen} durch systemtheoretische {Prozessanalyse}. \emph{Automatisierungstechnik}, 71(3). S.~209--218. DOI: {10.1515/auto-2022-0164}.
    }%
}
\vspace{2cm}

Cite as:

\vspace{0.1cm}
\noindent\fbox{%
    \parbox{\textwidth}{%
        R.~Graubohm, M.~Loba, M.~Nolte, and M.~Maurer, ``Identification of triggering conditions of {SOTIF} hazards through system-theoretic process analysis,'' (in German), \emph{Automatisierungstechnik}, vol.~71, no.~3, pp.~209--218, Mar.~2023, doi: {10.1515/auto-2022-0164}.
    }%
}
\vspace{2cm}

\end{@twocolumnfalse}
]
\noindent\begin{minipage}{\textwidth}
\hologo{BibTeX}:
\footnotesize
\begin{lstlisting}[frame=single]
@article{graubohm_identifikation_2023,
  author={{Graubohm}, Robert and {Loba}, Marvin and {Nolte}, Marcus and {Maurer}, Markus},
  title={Identifikation ausl{\"o}sender {Umst{\"a}nde} von {SOTIF}-{Gef{\"a}hrdungen} durch systemtheoretische {Prozessanalyse}},
  year={2023},
  journal = {at - Automatisierungstechnik},
  volume = {71},
  number = {3},
  pages = {209--218},
  month = mar,
  language = {de},
  doi = {10.1515/auto-2022-0164}
}
\end{lstlisting}
\end{minipage}

\clearpage

  \articletype{Methoden}

  \author*[1]{Robert Graubohm}
  \author[2]{Marvin Loba}
  \author[2]{Marcus Nolte} 
  \author[2]{Markus Maurer}
  \runningauthor{Graubohm et al.}
  \affil[1]{Technische Universität Braunschweig, Institut für Regelungstechnik (IfR), Braunschweig, Deutschland, E-Mail: graubohm@ifr.ing.tu-bs.de}
  \affil[2]{Technische Universität Braunschweig, Institut für Regelungstechnik (IfR), E-Mails: \{loba;nolte;maurer\}@ifr.ing.tu-bs.de}
  \title{Identifikation auslösender Umstände von SOTIF-Gefährdungen durch systemtheoretische Prozessanalyse}
  
  \runningtitle{Identifikation auslösender Umstände mittels STPA}
  \subtitle{Identification of triggering conditions of SOTIF hazards through System-Theoretic Process Analysis}
  
  \transabstract{Developers have to obtain a sound understanding of existing risk potentials already in the concept phase of driverless vehicles. Deductive as well as inductive SOTIF analyses of potential triggering conditions for hazardous behavior help to achieve this goal. In this regard, ISO~21448 suggests conducting a System-Theoretic Process Analysis (STPA). In this article, we introduce German terminology for SOTIF considerations and critically discuss STPA theory in the course of an example application, while also proposing methodological additions.}
\transkeywords{SOTIF, STPA, automated driving, hazard analysis}

  \abstract{Um bereits in der Konzeptphase autonomer Fahrzeuge einen fundierten Eindruck bestehender Risikopotenziale zu erhalten, werden im Zuge von deduktiven und induktiven SOTIF-Analysen mögliche auslösende Umstände für gefährliches Verhalten untersucht. In diesem Zusammenhang wird in der ISO~21448 die Durchführung einer systemtheoretischen Prozessanalyse (STPA) vorgeschlagen. In diesem Beitrag führen wir deutsche Terminologie für SOTIF-Betrachtungen ein und setzen uns im Zuge einer Anwendung kritisch mit der STPA-Theorie auseinander, wobei wir begleitend methodische Ergänzungen anregen.}

  \received{1. Dezember 2022}
  \accepted{1. Februar 2023}
  \journalname{at–Automatisierungstechnik}
  \journalyear{2023}
  \journalvolume{71}
  \journalissue{3}
  \startpage{209}
  \DOI{10.1515/auto-2022-0164}

\maketitle

\section{Einleitung} 

Die Norm ISO~21448~\cite{international_organization_for_standardization_iso_2022} beschäftigt sich mit den Residualrisiken des Einsatzes komplexer Assistenzsysteme und automatisierter Fahrfunktionen in Straßenfahrzeugen, die nicht unmittelbar Resultat von Ausfällen durch Fehler in Hardware und Software im Sinne der Funktionalen Sicherheit sind. Dieser Teil einer holistischen Sicherheitsbetrachtung für innovative Straßenfahrzeuge wird als \emph{SOTIF} (engl.~\foreignlanguage{english}{safety of the intended functionality}) beziehungsweise mit deutschem Titel als "`Sicherheit der beabsichtigten Funktionalität"' bezeichnet. Mit dem Ziel, bereits in der Konzeptphase einen fundierten Eindruck bestehender Risikopotenziale zu erhalten, beschreibt die 2022 erschienene ISO~21448 un­ter an­de­rem eine systematische Analyse möglicher auslösender Umstände (engl.~\foreignlanguage{english}{triggering conditions}) sowie derer Konsequenzen. Neben deduktiver Methodik, in der insbesondere funktionale Insuffizienzen auf Basis möglichen unsicheren Verhaltens in der angestrebten Betriebsumgebung identifiziert und analysiert werden, ist zusätzlich auch eine Analyse auslösender Umstände im Zuge eines induktiven Vorgehens empfohlen~\cite[Anhang~B.3.3]{international_organization_for_standardization_iso_2022}.

Ein in der ISO~21448 genannter und im Anhang der Norm auszugsweise illustrierter Ansatz zur systematischen Identifikation möglicher auslösender Umstände ist die Durchführung einer systemtheoretischen Prozessanalyse (STPA)~\cite[Anhang~B.4]{international_organization_for_standardization_iso_2022}. In diesem Beitrag überprüfen wir die Eignung des Ansatzes für die genannten Ziele anhand einer eigenen Durchführung der STPA für ein autonomes Fahrzeug. Neben der konkreten Darlegung unserer Ergebnisse dokumentieren wir dabei insbesondere vorgenommene Anpassungen unserer Anwendung im Vergleich zu den veröffentlichten allgemeinen Durchführungsanweisungen für die STPA~\cite{leveson_stpa_2018}.

In \autoref{sec:VerwArbeiten} werden verwandte Arbeiten zu beiden Aspekten~-- Durchführung einer STPA im Kontext des automatisierten Fahrens und Identifikation auslösender Umstände innerhalb von SOTIF-Betrachtungen~-- vorgestellt. Anschließend wird in \autoref{sec:Terminologie} auf die eine SOTIF-Gefährdungsanalyse begleitende Terminologie eingegangen. \autoref{sec:Prozessanalyse} enthält die Vorstellung unserer Strategie zur Durchführung einer STPA sowie wichtiger Resultate. Abschließend werden die Ergebnisse in \autoref{sec:Diskussion} diskutiert und in \autoref{sec:Fazit} ein Fazit gezogen.

\section{Verwandte Arbeiten} 
\label{sec:VerwArbeiten} 
Wie zuvor beschrieben untersuchen wir in diesem Beitrag die Möglichkeit, mithilfe des Ansatzes STPA auslösende Umstände im Sinne der SOTIF-Betrachtungen systematisch zu identifizieren. In der Literatur finden sich insbesondere für Teilaspekte dieser Fragestellung wichtige Vorarbeiten, die in unsere Vorgehensstrategie (s.~\autoref{ssec:Vorgehensstrategie}) eingeflossen sind. Die zentralen Quellen für die Grundlagen der STPA stellen die Vorstellung der Methode durch \textcite{leveson_engineering_2012} sowie das STPA-Handbuch von \textcite{leveson_stpa_2018} dar.

Literaturquellen mit Anwendungen der STPA im Kontext des automatisierten Fahrens liefern insbesondere Vorbilder für Kontrollstrukturen, denen innerhalb der Methode STPA große Bedeutung zukommt. Beispiele sind die Modelle von \textcite{bagschik_safety_2017} für ein automatisiert fahrendes Absicherungsfahrzeug und von \textcite{stolte_safety_2016} für Aktuatorsysteme. 
Die Arbeiten von \textcite{schnieder_generische_2020,schnieder_leitfaden_2020} liefern weitere Anwendungs- und Modellierungsbeispiele für die STPA im Kontext des automatisierten Fahrens. Im Gegensatz zum Fokus unserer Analyse konzentrieren sich \citeauthor{schnieder_leitfaden_2020} in ihren Arbeiten allerdings auf konventionelle Straßenfahrzeuge mit Autopilotfunktionalität.

Als Teil des Forschungsprojekts PEGASUS stellen \textcite{bode_identifikation_2019} in einem Technischen Bericht eine Methode zur Identifikation und Quantifizierung von Automatisierungsrisiken vor. Im Zuge dessen untersuchen sie auch die Anwendbarkeit des STPA-Ansatzes und übernehmen einzelne Schlüsselworte aus der STPA-Theorie für ihre Methode. Bei der detaillierten Auswertung ihrer schlüsselwortbasierten Gefährdungsanalyse gehen \citeauthor{bode_identifikation_2019} auch auf "`auslösende Umgebungsbedingungen"' ein und weisen auf die Anknüpfungspunkte mit dem SOTIF-Prozess hin. Der wichtigste Unterschied zu unserer Analyse ist die Zielsetzung der Autor*innen, Automatisierungsrisiken und entsprechende Auslöser zu finden, statt zielgerichtet potenzielle auslösende Umstände zu identifizieren.

Die SAE~J3187~\cite{functional_safety_committee_sae_2022} gibt Empfehlungen bezüglich der Anwendung einer STPA für sicherheitskritische Fahrzeugsysteme. Dabei wird spezifisch auf die Nutzung der STPA-Methode für SOTIF-Aktivitäten eingegangen und eine Fallstudie für ein automatisiertes Niedriggeschwindigkeitsfahrzeug präsentiert. Die Verwandtschaft des Vorgangs der Identifikation auslösender Umstände mit dem Schritt der Beschreibung von Verlustszenarien im Zuge einer STPA wird in der SAE~J3187 angeführt~\cite[Abschn.~7.1]{functional_safety_committee_sae_2022}, aber nicht näher beleuchtet.

Im Kontext der STPA-Theorie ist äußerst wichtig festzuhalten, dass die Methode keinen wesentlichen Beitrag zu einer~-- initial notwendigen~-- Ge\-fähr\-dungs\-iden\-ti\-fi\-ka\-tion leistet. Vielmehr wird im ersten Schritt einer STPA gefordert, mögliche Gefährdungen, die vom untersuchten System ausgehen, als Startpunkt der Analyse niederzuschreiben. Für etablierte sicherheitskritische Systeme ist vorstellbar, dass sämtliche relevante Gefährdungen bereits bekannt sind. In einem innovativen Feld wie dem automatisierten Fahren sind Ansätze zur systematischen Gefährdungsidentifikation allerdings weiterhin Inhalt von Forschungsarbeiten~\cite{graubohm_towards_2020}. Der Beitrag einer STPA ist vielmehr die Identifikation von Ursachen und Konkretisierung vorliegender Gefährdungen durch Beschreibung kausaler Zusammenhänge anhand von Szenarien~-- die Zielsetzung ähnelt also der einer Fehlerbaumanalyse. Diese Tatsache wird in der zuvor angeführten Fachliteratur zur STPA (insbesondere~\cite{leveson_engineering_2012,leveson_stpa_2018}) grundsätzlich erläutert, jedoch in Publikationen über die Anwendung von Methoden der Gefährdungsanalyse häufig anders dargestellt. Beispiele für aktuelle Veröffentlichungen, die in ihren Ausführungen eindeutig suggerieren, STPA könne im Kontext des automatisierten Fahrens als Ansatz für die Gefährdungsidentifikation eingesetzt werden, sind~\cite{becker_safety_2020,capito_methodology_2021,jianyu_model-based_2021,khatun_approach_2021}.

Die SOTIF-Norm ISO~21448 empfiehlt die Anwendung systematischer Ansätze zur Identifikation möglicher auslösender Umstände~\cite[Abschn.~7.3]{international_organization_for_standardization_iso_2022}. Die Methoden beruhen dabei entweder auf der Analyse möglicher Szenarien, in denen bekannte systemische Insuffizienzen zu Gefährdungen führen, oder der Untersuchung der Bedingungen einer Einsatzumgebung. Die Anwendung einer STPA ist entsprechend nur eine von vielen Möglichkeiten zur systematischen Identifikation auslösender Umstände (vgl.~\cite[Tabelle~4]{ international_organization_for_standardization_iso_2022}).

Mit dem Ziel, im Zuge von SOTIF-Betrachtungen mögliche auslösende Umstände zu ermitteln, präsentieren \textcite{zhu_systematization_2022} einen Ansatz, der auf einer zweidimensionalen heuristischen Matrix beruht. Die Autor*innen diskutieren dabei auch die Durchführung einer STPA als mögliche Alternative. Als Kritikpunkte an der STPA werden vorrangig die fehlenden Handlungsempfehlungen bei der Modellierung einer Kontrollstruktur und die fehlende Struktur bei der Ableitung von Szenarien und Ursachen in den einzelnen Teilschritten der Methode vorgetragen. Diese Defizite können wir grundsätzlich bestätigen, argumentieren allerdings, dass sie eine Anwendung nur erschweren und nicht ausschließen.

\section{Terminologie} 
\label{sec:Terminologie} 

\begin{figure*}[thpb]
      \centering
      \includegraphics[width=\textwidth]{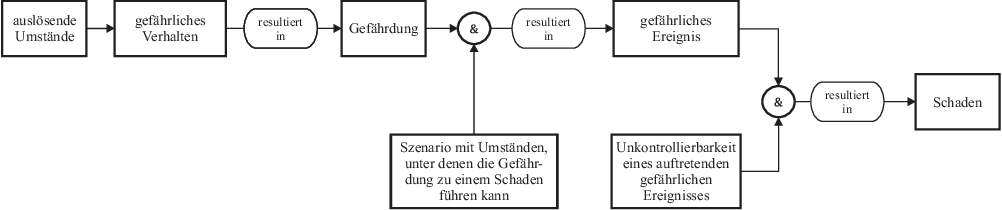}
      \caption{Visualisierung der Gefährdungsterminologie der ISO~21448 in Anlehnung an~\cite[S.~11]{international_organization_for_standardization_iso_2022}}
      \label{fig:Hazardmodell}
   \end{figure*}

Um die Gefährdungsanalyse unter Zuhilfenahme der STPA möglichst eng in Verbindung mit den Anforderungen der ISO~21448 zu bringen, haben wir die SOTIF-Terminologie rund um \emph{gefährliche Ereignisse} (engl.~\foreignlanguage{english}{hazardous events}) analysiert und teilweise konkretisiert. \autoref{fig:Hazardmodell} führt die wesentlichen Begriffe auf und visualisiert ihren Zusammenhang im Sinne der SOTIF-Norm.

Die Abfolge beginnt mit auslösenden Umständen. Ein auslösender Umstand wird durch die ISO~21448 als Bestandteil eines Szenarios definiert, der Initiator für gefährliches Verhalten ist~\cite[Abschn.~3.30]{international_organization_for_standardization_iso_2022}. Dieses \emph{gefährliche Verhalten} wird in der Norm von \emph{Gefährdungen} unterschieden. Hierbei ist besonders zu erwähnen, dass die ISO~21448 keine Definition für das im Normtext vielfach angeführte Konzept \emph{gefährliches Verhalten} (engl.~\foreignlanguage{english}{hazardous behaviour}) anbietet. Unter Zuhilfenahme der Ausführungen und Anmerkungen in der SOTIF-Norm definieren wir \emph{gefährliches Verhalten} für unsere Arbeit folgendermaßen: 
\begin{quote}
\normalsize Konkrete Aktion des Fahrzeugs, die mindestens eine Gefährdung nach sich zieht, mit zeitlicher Entwicklung~-- extern als Bewegung beziehungsweise dynamisches Verhalten wahrnehmbar.
\end{quote}
Darüber hinaus unterscheidet die ISO~21448 zwischen der \emph{Gefährdung} (der potenziellen Quelle von Schaden hervorgerufen durch gefährliches Verhalten~\cite[Abschn.~3.11]{international_organization_for_standardization_iso_2022}) und dem \emph{gefährlichen Ereignis} (engl.~\foreignlanguage{english}{hazardous event}). In \autoref{fig:Hazardmodell} wird deutlich, dass Szenarienelemente wie potenziell betroffene Personen oder Objekte erst in die Beschreibung des gefährlichen Ereignisses einfließen. Dadurch gibt es einen konkreten Unterschied zwischen dem Gefährdungsbegriff der SOTIF-Norm und beispielweise \textcite[Abschn.~26.3]{ericson_hazard_2005}, der für das Vorhandensein einer Gefährdung auch immer ein potenziell betroffenes Objekt voraussetzt.

Die Übertragung der SOTIF-Terminologie auf den methodischen Rahmen einer STPA ist ohne Weiteres möglich und kann grundsätzlich in Referenz zur STPA-Beispielanwendung der ISO~21448~\cite[Anhang~B.4]{international_organization_for_standardization_iso_2022} erfolgen. Wir teilen allerdings nicht die Auffassung des Anwendungsbeispiels, dass der STPA-Begriff \emph{unsichere Kontrollaktionen} auch mit dem SOTIF-Begriff \emph{gefährliches Verhalten} betitelt werden kann. Auch \textcite[Abschn.~5.1.2]{becker_safety_2020} geben an, unsichere Kontrollaktionen seien analog zu potenziell gefährlichem Verhalten im SOTIF-Kontext. Zumindest gemäß unserer zuvor aufgeführten Definition ist \emph{gefährliches Verhalten} die konkrete Fahrzeugbewegung, während \emph{unsichere Kontrollaktionen} in der STPA-Theorie im Wesentlichen Fehler an der Schnittstelle eines Reglers zu anderen Reglern oder Aktuatoren in der betrachteten Kontrollstruktur sind. Insofern stellen die Begriffe in unserer Untersuchung keine Synonyme dar.

Über die SOTIF-Terminologie hinaus spielen \emph{Verluste} (engl.~\foreignlanguage{english}{losses}) in der STPA-Theorie eine wichtige Rolle. Im Zusammenhang mit einer STPA geht es dabei konkret um den Verlust von etwas, das für Menschen von Wert ist~\cite[S.~16]{leveson_stpa_2018}. Neben dem Verlust menschlichen Lebens und Verletzungen sind also auch andere~-- zum Beispiel rein wirtschaftliche~-- Schäden eingeschlossen. Der Begriff \emph{Verlust} ähnelt damit dem \emph{Schaden} (engl.~\foreignlanguage{english}{harm}) gemäß ISO/IEC~GUIDE~51~\cite[Abschn.~3.1]{international_organization_for_standardization_isoiec_2014}.

\section{Systemtheoretische Prozessanalyse} 
\label{sec:Prozessanalyse} 

\subsection{Vorgehensstrategie}
\label{ssec:Vorgehensstrategie}
Auf Basis der zuvor angeführten Literatur und in Referenz zur Beispielanwendung einer STPA im Anhang der ISO~21448 haben wir eine Vorgehensstrategie erarbeitet, die erlaubt, aus definierten Szenarien unter Zuhilfenahme der STPA-Methode auslösende Umstände zu identifizieren. Konkret wurde die allgemeine STPA-Prozesskette übernommen und um die nachgelagerte explizite Beschreibung von auslösenden Umständen (als Gefährdungsursachen in Verlustszenarien) ergänzt: 
\begin{enumerate}
\item Identifiziere Verluste und Gefährdungen
\item Modelliere die Kontrollstruktur
\item Identifiziere unsichere Kontrollaktionen
\item Identifiziere Verlustszenarien
\item Beschreibe auslösende Umstände
\end{enumerate}
Den initial notwendigen Prozessschritt, Verluste und Gefährdungen aufzuführen, haben wir bereits zuvor erwähnt. Die Modellierung der Kontrollstruktur (Schritt~2) wird in \autoref{ssec:Kontrollstruktur} anhand unseres Anwendungsbeispiels näher beleuchtet. In Schritt 3 wird zur Beschreibung unsicherer Kontrollaktionen auf Basis der dokumentierten Gefährdungen und Kontrollstruktur des betrachteten Systems ein Katalog von Leitworten eingesetzt, der durch die STPA-Methodik vorgegeben ist (vgl.~\cite[S.~34ff]{leveson_stpa_2018}):
\begin{itemize}
\item Keine Bereitstellung
\item Falsche Bereitstellung
\item Zu frühe oder zu späte Bereitstellung
\item Zu lange oder zu kurze Bereitstellung
\end{itemize}
Ähnlich wie in der ISO~21448 vorgeschlagen, führen wir zusammen mit unsicheren Kontrollaktionen auch drohende Konsequenzen~-- also das initiierte gefährliche Verhalten~-- auf, um die Verknüpfung mit den Gefährdungen zu stärken. Inspiriert durch verwandte Arbeiten sehen wir an der Schnittstelle der Verfahrensschritte~3 und~4 darüber hinaus den Einsatz der Inhalte eines Modells allgemeiner Kausalitäten (s.~\cite[Abschn.~8.4.10]{leveson_engineering_2012} und~\cite[Anhang~G]{leveson_stpa_2018}) vor. Auf dieser Basis beschreiben wir konkrete Verlustszenarien. Dieser Detaillierungsschritt fehlt in der Beispielanwendung der ISO~21448, da dort nur Kausalfaktoren spezifiziert~\cite[Anhang~B.4.5]{international_organization_for_standardization_iso_2022} werden. Diese Abweichung von der STPA-Theorie könnte nach unserer Ansicht zu einer deutlichen Einschränkung der identifizierbaren auslösenden Umstände führen, da nur bekannte Insuffizienzbedingungen betrachtet werden.

Die STPA-Literatur lässt verschiedene Detaillierungsgrade für Verlustszenarien zu. Mindestens sollten der die unsichere Kontrollaktion initiierende Regler und die Konsequenz im Sinne einer Gefährdung verdeutlicht werden~\cite[S.~42-53]{leveson_stpa_2018}. Wir haben uns für eine etwas ausführlichere Beschreibung von Szenarien entschieden, die zusätzliche Informationen über Kausalzusammenhänge enthält~-- also beispielsweise einen Sensor spezifiziert, der dem Regler falsche Daten übermittelt und dadurch eine Gefährdung hervorruft. Grundsätzlich könnte die Verlustszenarienbeschreibung einer STPA auch bereits Ursachen in der Systemumgebung vollständig angeben, was für die systematische Analyse möglicher auslösender Umstände im nächsten Schritt unserer Durchführung allerdings nicht sinnvoll wäre.

Abschließend werden in unserer Durchführung die Verlustszenarien hinsichtlich möglicher auslösender Umstände analysiert. Hierbei ist zu beachten, dass Szenarien eine Vielzahl möglicher auslösender Umstände erlauben können~-- einzelne Umstände wiederum aber auch eine Vielzahl möglicher gefährlicher Ereignisse hervorrufen können. Entsprechend wird in der ISO~21448 die Wichtigkeit einer Rückverfolgbarkeit betont~\cite[Abschn.~7.3.1]{international_organization_for_standardization_iso_2022}. Eine geeignet dokumentierte STPA kann unserer Ansicht nach hierbei wichtige Unterstützung leisten, da Verknüpfungen systematisch betrachtet werden. Entsprechend haben wir entschieden, im letzten Verfahrensschritts auch funktionale Insuffizienzen zu spezifizieren, die die Kausalität zwischen Verlustszenarien und auslösenden Umständen begründen.

\begin{figure}[thpb]
      \centering
      \includegraphics[width=\linewidth]{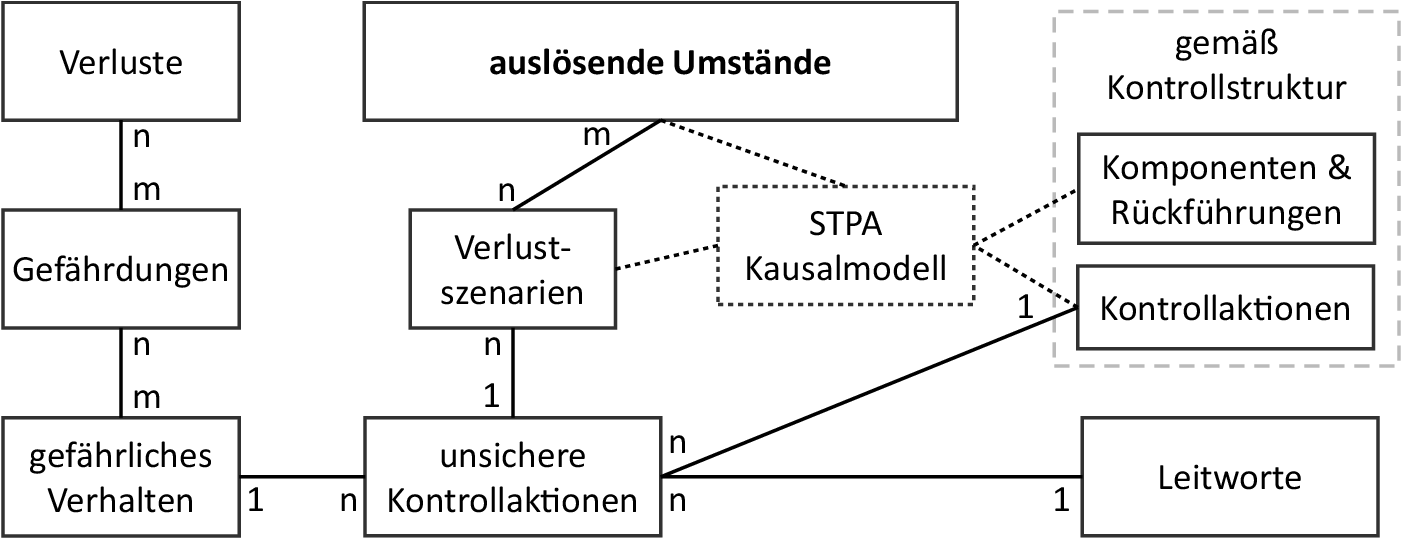}
      \caption{Datenstruktur für die Dokumentation der systemtheoretischen Prozessanalyse}
      \label{fig:Datenstruktur}
   \end{figure}

Der Zusammenschluss \emph{\citeauthor{mit_partnership_for_systems_approaches_to_safety_and_security_psass_stamp_2020}} pflegt eine Liste verfügbarer Softwarewerkzeuge für die Unterstützung der Durchführung einer STPA~\cite{mit_partnership_for_systems_approaches_to_safety_and_security_psass_stamp_2020}. Die im Zuge unserer Arbeiten untersuchten Werkzeuge \emph{STAMP Workbench}\footnote{\href{https://www.ipa.go.jp/english/sec/complex\_systems/stamp.html}{https://www.ipa.go.jp/english/sec/complex\_systems/stamp.html}} und \emph{XSTAMPP}\footnote{\href{https://github.com/SE-Stuttgart/XSTAMPP}{https://github.com/SE-Stuttgart/XSTAMPP}} waren nicht in der Lage, die hier geschilderte Vorgehensstrategie geeignet zu begleiten und zu dokumentieren. Insbesondere war die Möglichkeit, Verlustszenarien zu beschreiben und mit anderen Prozessartefakten zu verknüpfen, stark eingeschränkt. Entsprechend haben wir uns für eine eigene Dokumentation der Gefährdungsanalyse in Tabellenform entschieden. Die hierbei implementierte Datenstruktur inklusive Relationen ist in \autoref{fig:Datenstruktur} gezeigt. 

Verluste, Gefährdungen und gefährliches Verhalten können aufgeführt und in beliebigem Verhältnis miteinander verknüpft werden. Unsichere Kontrollaktionen werden mithilfe der spezifisch modellierten Kontrollstruktur aus genau einem Leitwort gemäß STPA-Theorie gebildet und enthalten genau ein gefährliches Verhalten. Verlustszenarien werden dann nach STPA-Kausalmodell generiert und inkludieren genau eine unsichere Kontrollaktion. Schließlich können aus den Verlustszenarien beliebig viele auslösende Umstände abgeleitet werden, die jeweils auch mehrere Verlustszenarien begründen können.

\subsection{Kontrollstruktur eines automatisierten Fahrzeugs} 
\label{ssec:Kontrollstruktur} 

\begin{figure*}[thpb]
      \centering
      \includegraphics[width=\textwidth]{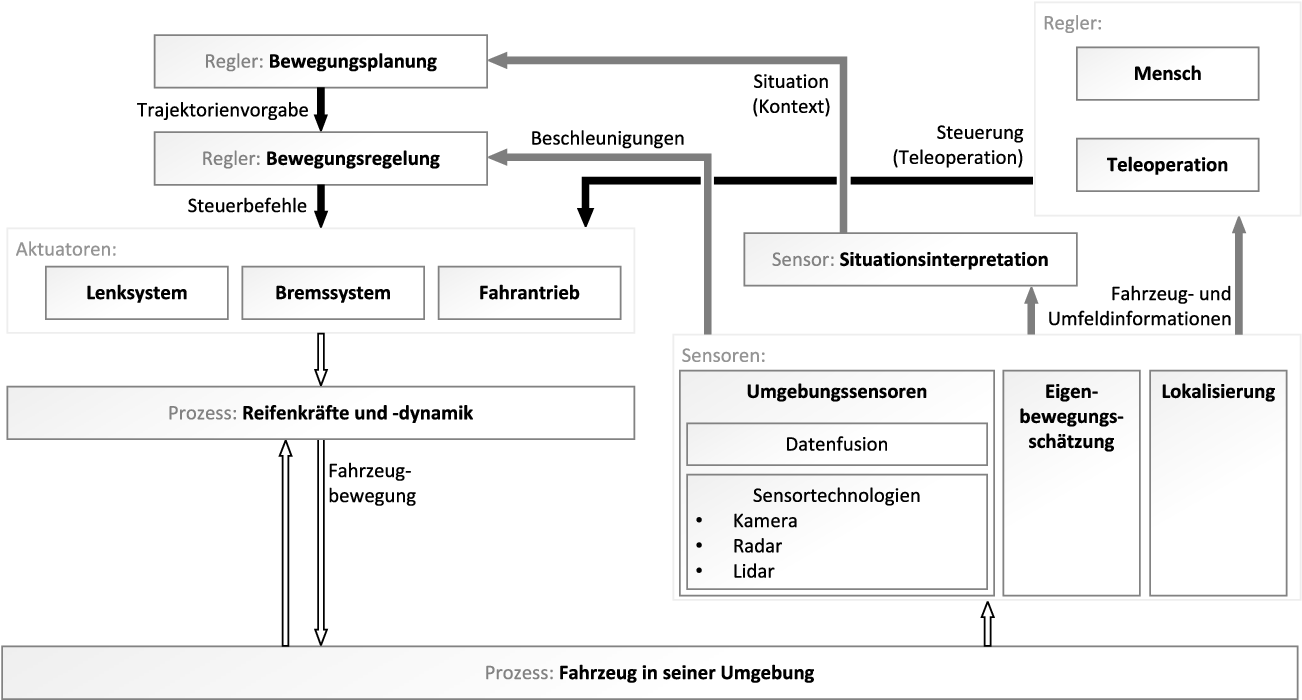}
      \caption{Allgemeine Kontrollstruktur eines automatisierten Fahrzeugs (schwarze Pfeile zeigen Kontrollaktionen, graue Pfeile zeigen Rückführungen und weiße Pfeile sonstige Verknüpfungen)}
      \label{fig:Kontrollstruktur}
   \end{figure*}

\autoref{fig:Kontrollstruktur} zeigt die für die Untersuchung erzeugte STPA-Kontrollstruktur bestehend aus Prozessen, Sensoren, Reglern und Aktuatoren. Anwendungsfall ist ein Fahrzeug der Automatisierungsstufe SAE~Level~4~\cite{on-road_automated_driving_orad_committee_sae_2021}. Nach dem vereinfachten Modell der Bundesanstalt für Straßenwesen~\cite{bundesanstalt_fur_strasenwesen_selbstfahrende_2021} handelt es sich also um den \emph{autonomen Modus}.

Wir haben uns für eine eng an der STPA-Theorie orientierte konsequente Unterscheidung zwischen den Komponenten Prozess, Sensor, Regler und Aktuator innerhalb der Kontrollstruktur entschieden (vgl.~\cite[Abbildungen~8.1, 8.6 und 8.8]{leveson_engineering_2012}), die sich deutlich vom sehr abstrakten Kontrollstrukturvorschlag der Beispielanwendung in der ISO~21448~\cite[Abbildung~B.7]{international_organization_for_standardization_iso_2022} unterscheidet. Wegen des besonderen Charakters der Absicherung einer Teleoperation haben wir menschliche Steuerungsfehler und ihre Ursachen als Quelle von Gefährdungen in unserer weiteren Beispielanalyse nicht näher untersucht, sondern uns auf auslösende Umstände im Sinne von Umgebungsbedingungen des Fahrzeugs konzentriert. Grundsätzlich ist unserer Kontrollstruktur aber zu entnehmen, dass die Möglichkeit einer Teleoperation eine wichtige Quelle für gefährliches Verhalten aufgrund unsicherer Kontrollaktionen darstellt.

\subsection{Durchführung und Identifikation auslösender Umstände} 
\label{ssec:Identifikation} 

Entsprechend der in \autoref{ssec:Vorgehensstrategie} vorgestellten Prozessfolge ist die Identifikation von Verlusten und Gefährdungen im jeweiligen Anwendungsfall der erste Schritt der Methode. Verwandte STPA-Anwendungsbeispiele aus der Literatur weisen vorwiegend sehr abstrakte Gefährdungsbeschreibungen auf, die in unserem Fall aber sowohl Ausführung als auch Bewertung des Nutzens erschwert hätten. Ein Beispiel für eine derart abstrakte Gefährdung ist "`Autopilot führt ein unsicheres Fahrmanöver aus"'~\cite[S.~22]{schnieder_leitfaden_2020}. Für die Analyse der Eignung des STPA-Ansatzes zur systematischen Identifikation von auslösenden Umständen beschränken wir uns auf ein konkretes Gefährdungsbeispiel, das innerhalb der STPA-Theorie klar mit dem Verlust "`Verlust von Menschenleben oder Verletzung von Menschen"' verknüpft ist:
\begin{quote}
\normalsize Unterschreitung eines angemessenen Mindestabstandes zu Fußgänger*innen.
\end{quote}
Damit wählen wir eine Gefährdung von äußerster Wichtigkeit im Kontext des automatisierten Fahrens~-- insbesondere in städtischer Umgebung. Der Verlust und die Gefährdung sind darüber hinaus absehbar mit diversen SOTIF-relevanten möglichen auslösenden Umständen verknüpft. Für andere Ausprägungen von Verlusten aus der STPA-Literatur, die nicht eng mit menschlichem Schaden verknüpft sind~-- zum Beispiel die Abnahme von Kundenzufriedenheit oder negative Umwelteinflüsse~--, ist dies eher fraglich. 

Im Übergang zu den weiteren Verfahrensschritten mit dem abschließenden Ziel der Identifikation von auslösenden Umständen wird dann auf Basis der untersuchten Gefährdung mögliches gefährliches Verhalten konkretisiert. Wir unterscheiden hierfür zwei mögliche Ausprägungen von Fahrzeugverhalten, die im Straßenverkehr zu Unterschreitungen einer sicheren Distanz zu Fußgänger*innen führen können:
\begin{itemize}
\item Ego-Fahrzeug bremst nicht bis zum Stillstand ab (bei Zusammentreffen mit einem Fußgänger oder einer Fußgängerin im eigenen Fahrstreifen).
\item Ego-Fahrzeug ändert seinen Kurswinkel und verlässt die Fahrstreifengrenzen (in Richtung einer Fußgängerin oder eines Fußgängers in der Nähe).
\end{itemize}
Dieser Analyseschritt ist im Kontext einer~-- an der SOTIF-Norm orientierten~-- STPA für ein automatisches Fahrzeugführungssystem äußerst kritisch zu betrachten: Wird gefährliches Verhalten vergessen, würden auch entsprechende unsichere Kontrollaktionen, die genau dieses Verhalten verursachen könnten, nicht mehr dokumentiert. Das hat unmittelbare Konsequenzen für die Menge erzeugter Verlustszenarien, die als Ursachenanalyse für gegebene Gefährdungen im Kern das Resultat einer STPA darstellen. In verwandten Arbeiten erfolgt dieser Zwischenschritt der dokumentierten Überlegung möglicher gefährlicher Verhaltensausprägungen durch das untersuchte System meist nicht einmal. Dadurch liegt dann noch größere Verantwortung als ohnehin schon in dem Verfahrensschritt 3 einer STPA, da unsichere Kontrollaktionen direkt aus der Gefährdung und in Referenz zur Kontrollstruktur abgeleitet werden müssen.

Mithilfe der im \autoref{ssec:Vorgehensstrategie} aufgeführten Leitworte und der vorgestellten exemplarischen Kontrollstruktur können aus den Kontrollaktionen \emph{Trajektorienvorgabe}, \emph{Steuerbefehle} und \emph{Steuerung (Teleoperation)} mögliche unsichere Kontrollaktionen generiert werden. Wir identifizieren 14 unsichere Kontrollaktionen mit direktem Bezug zu den beiden oben beschriebenen Ausprägungen von gefährlichem Verhalten. Von diesen 14 unsicheren Kontrollaktionen betrachten wir im Folgenden zwölf, da zwei Ausprägungen mit menschlichen Fehlern im Zuge möglicher Teleoperation zusammenhängen, die einer gesonderten Analyse bedürfen. Ein Beispiel für eine unsichere Kontrollaktion in unserem Anwendungsfall ist:
\begin{quote}
\normalsize Der Bewegungsregler gibt keinen Bremsbefehl, sodass das Ego-Fahrzeug nicht bis zum Stillstand abbremst (bei Zusammentreffen mit einem Fußgänger oder einer Fußgängerin im eigenen Fahrstreifen).
\end{quote}

Anschließend kann das STPA-Kausalmodell~\cite[Anhang~G]{leveson_stpa_2018} eingesetzt werden, um Verlustszenarien zu spezifizieren. Hierbei wird in Referenz zu~\cite[S.~45]{leveson_stpa_2018} das Kausalmodell zusätzlich um physische Ausfälle von Reglern ergänzt, die das funktionale Kausalmodell nicht aufführt. Darüber hinaus wird eine Fallunterscheidung für die Annäherung an eine*n Fußgänger*in im eigenen Fahrstreifen ohne rechtzeitige und ausreichende Bremsung eingeführt: Die Person kann sich entweder bereits über einen längeren Zeitraum im Fahrstreifen befunden haben oder unmittelbar in den Fahrkorridor des heranfahrenden Fahrzeugs eindringen. Insgesamt werden dadurch auf Basis von zwei Möglichkeiten gefährlichen Verhaltens und zwölf unsicheren Kontrollaktionen 103 Verlustszenarien beschrieben. Die Vielzahl der Szenarien resultiert dabei aus der in einer STPA vorgenommenen Unterscheidung verschiedener Kausaltypen in der Kontrollstruktur. Ein Beispiel für ein Verlustszenario im Kontext der oben wiedergegebenen unsicheren Kontrollaktion ist:
\begin{quote}
\normalsize Das Ego-Fahrzeug trifft während der automatischen Fahrzeugführung auf eine*n Fußgänger*in im eigenen Fahrstreifen. Der Bewegungsregler gibt keinen Bremsbefehl, da die Eigenbewegungsschätzung unzureichend funktioniert. Ego-Fahrzeug bremst nicht bis zum Stillstand ab, bevor es die Person erreicht. Die fehlende Abbremsung führt zu einer Kollision zwischen Ego-Fahrzeug und der Person.
\end{quote}
Allerdings ist unmittelbar eine Reduktion der 103 Szenarien auf 55 Szenarien möglich. Nur diese 55 Szenarien beinhalten eine Kausalität, die in potenziellem Zusammenhang mit auslösenden Umständen gemäß SOTIF steht. Herausgefilterte Ursachen sind dabei insbesondere Fehler und Ausfälle in physischen Steuergeräten oder Kommunikationsnetzen. Die genannten Fehlertypen sind offensichtlich dem Bereich der Funktionalen Sicherheit zuzuordnen (vgl. ISO~26262~\cite{international_organization_for_standardization_iso_2018}) und keine funktionalen Insuffizienzen im Sinne einer SOTIF-Analyse. Während der Auswahl SOTIF-relevanter Szenarien wurde außerdem festgestellt, dass für Steuergeräte im Anwendungsfall die in einer STPA für Regler vorgesehene Unterscheidung in "`Fehler eines Kontrollalgorithmus"' und "`Defizite eines Prozessmodells"' unnötig scheint. Zwar begründen beide Ausprägungen teilweise auch SOTIF-relevante Szenarien, mögliche Ursachen sind aber nicht zu unterscheiden. Eine Reduktion des in der STPA-Theorie vorgesehenen Detaillierungsgrads könnte an dieser Stelle also die Effizienz steigern.

Für die abschließende Beschreibung möglicher auslösender Umstände stellt sich in Hinblick auf die Kontrollstruktur die Frage der Verortung. Wie in \autoref{sec:Terminologie} angeführt, werden auslösende Umstände als Szenarienbestandteil definiert, was dafür spricht, sie als Teil des Prozessblocks "`Fahrzeug in seiner Umgebung"' zu sehen. Gegebenenfalls kann allerdings vorhersehbare Fehlbenutzung durch Menschen, welche eine Unterkategorie auslösender Umstände ist, auch an anderer Stelle geschehen (zum Beispiel durch direkte Steuerbefehle). Die Auswirkungen (funktionale Insuffizienzen aufgrund eines Auslösers) können in jedem Fall an beliebiger Stelle innerhalb der Kontrollstruktur liegen. Die ISO~21448~\cite[Abschn.~3.30, Anmerkung~2]{international_organization_for_standardization_iso_2022} liefert hierfür ein Beispiel, in dem ein automatisches Notbremssystem aufgrund fehlerhafter Klassifikation auf ein Verkehrsschild bremst. Als auslösender Umstand wird im Beispiel das Auftreten eines derartig problematischen Schildes benannt~-- also ein Ereignis im Umfeld des Fahrzeugs. Die beschriebene Insuffizienz liegt aber in einer Kontrollstruktur am Ort der Klassifikation.

Mit unserer Konzentration auf externe Ereignisse beschreiben wir für die 55 näher betrachteten Verlustszenarien auslösende Umstände ausgehend vom Prozessblock "`Fahrzeug in seiner Umgebung"'. Die Identifikation auslösender Umstände hat das primäre Ziel, ein in den Szenarien beschriebenes gefährliches Verhalten im Sinne der über die STPA herausgearbeiteten systemischen Ursachen zu erklären. Wir beschränken uns hierbei auf 18 Ausprägungen von auslösenden Umständen, die wesentliche Dimensionen abdecken\footnote{Beispielsweise schlägt die ISO~21448 vor, zwischen regnerischem Wetter, Regen, Schneeregen, angesammeltem Schnee, Schneefall, Hagel, Nebel sowie diversen Straßen- und Fahrzeugbeeinflussungen zu differenzieren~\cite[Anhang~B]{international_organization_for_standardization_iso_2022}. Für die Überprüfung der generellen Eignung unserer Vorgehensstrategie ist es aber ausreichend, die auslösenden Umstände Regenfall, regennasse Straße, verschmutzte Straße und verschmutztes Fahrzeug zu unterscheiden.}. Einzelne auslösende Umstände sind dabei mit über 40 Szenarien verknüpft, während gleichzeitig für einzelne Verlustszenarien bis zu 14 mögliche auslösende Umstände möglich sind.

Für die Dokumentation und Erläuterung der Verknüpfung von Verlustszenarien aus der STPA-Anwendung mit möglichen auslösenden Umständen wird an der Schnittstelle jeweils eine funktionale Insuffizienz spezifiziert. Das zu späte Reagieren auf eine Person im Fahrbereich eines automatisierten Fahrzeugs kann zum Beispiel innerhalb einer STPA auf falsche Informationen aus dem Wahrnehmungssystem zurückgeführt werden. Als möglichen auslösenden Umstand wird die tiefstehende Sonne dokumentiert. Die Ausführungen zur verknüpfenden funktionalen Insuffizienz erklären dann, dass die tiefstehende Sonne Sensoren eines Wahrnehmungssystems blenden kann und dass das Wahrnehmungssystem aufgrund der eingesetzten Sensortechnologie dagegen unzureichend robust sein könnte.

\section{Diskussion} 
\label{sec:Diskussion} 

Das grundlegende Ziel unserer vorgestellten Arbeit ist eine Untersuchung und Bewertung der Eignung einer STPA, um eine belastbare Datenbasis für auslösende Umstände zu schaffen. Grundsätzlich stellen wir eine große Verwandtschaft zwischen der systemischen Ursachenanalyse einer STPA mit einer systematischen Identifikation von auslösenden Umständen im SOTIF-Prozess fest. Diese wird auch bereits im STPA-Beispiel im Anhang der ISO~21448 deutlich. Eine direkt anhand Literatur durchgeführte STPA für den Anwendungsfall einer Fahrzeugführungsfunktion eines automatisierten Fahrzeugs resultiert aber nicht unbedingt in einer Liste von auslösenden Umständen: Die Vorgaben im STPA-Verfahrensschritt~4 (Identifikation von Verlustszenarien) sind dafür zu vage. Wir schlussfolgern also, dass mit dem Ziel, eine SOTIF-Analyse zu unterstützen, Prozessvorgaben für die Anwendung der STPA konkretisiert und gegebenenfalls erweitert werden sollten, wie in \autoref{ssec:Vorgehensstrategie} erläutert.

Ein direkter Abgleich unserer Ergebnisse mit der Beispielanwendung der STPA im Anhang der ISO~21448 ist nicht ohne Weiteres möglich, da in der Norm Analyseergebnisse nur auszugsweise wiedergegeben sind. Darüber hinaus endet die Ursachenanalyse je Szenario bei sogenannten Kausalfaktoren, bei denen es sich neben auslösenden Umständen auch um funktionale Insuffizienzen oder Ausgabeinsuffizienzen handeln kann~\cite[Anhang~B.4.5]{international_organization_for_standardization_iso_2022}. Eine Zuordnung, welche Kausalfaktoren tatsächlich auslösende Umstände beschreiben, findet nicht statt.

Wie bereits angeführt stellen wir außerdem fest, dass über die Methode STPA hinaus im Zuge eines deduktiven Vorgehens zur Identifikation möglicher auslösender Umstände die Aufführung von \emph{gefährlichem Verhalten} ein ausgesprochen kritischer Verfahrensschritt ist. \emph{Gefährliches Verhalten} schließt die Lücke zwischen~-- bestenfalls anhand etablierter Prozesse belastbar identifizierter~-- Gefährdungen und der Untersuchung konkreter funktionaler Insuffizienzen innerhalb einer SOTIF-Analyse. Wird ein mögliches, gegebenenfalls nur in bestimmten Szenarien relevantes gefährliches Fahrzeugverhalten übersehen, kann die gesamte folgende systematische Untersuchung von fehlenden Ausprägungen der Gefährdung gekennzeichnet sein. Das ist auch für eine STPA der Fall: Eine "`Vollständigkeit"' der unsicheren Kontrollaktionen hinsichtlich einer Gefährdung~-- und damit das Potenzial, alle auslösenden Umstände zu finden~-- hängt im untersuchten Prozess im Wesentlichen von der differenzierten Beschreibung der Ausprägungen gefährlichen Verhaltens ab.

In der ISO~21448 wird vorrangig eine Vorgehensweise skizziert, bei der anhand bekannter Ursachen auf resultierende Fahrzeugeffekte rückgeschlossen wird. Dieses Vorgehen wird im Anhang der SOTIF-Norm als "`induktive SOTIF-Analyse"' bezeichnet~\cite[Anhang~B.3.3]{international_organization_for_standardization_iso_2022}. Da hierbei gefährliches Verhalten systematisch hergeleitet wird, tritt die zuvor beschriebene Problematik nicht auf. Eine solche induktive Analyse geht allerdings von bekannten auslösenden Umständen aus und unterstützt deren systematische Identifikation kaum. Sie stellt entsprechend keine Alternative zum von uns untersuchten Vorgehen dar.

Grundsätzlich wird die Qualität der Ergebnisse einer Prozessanalyse durch ein umfangreiches Systemverständnis der Durchführenden erhöht. \textcite{zhu_systematization_2022} kritisieren aber zusätzlich, dass die fehlenden Handlungsempfehlungen für die Durchführung einer STPA zu zufallsbedingtem Brainstorming führen. Wir stellen hingegen fest, dass die Verfahrensschritte zur Ursachenanalyse auf Basis von unsicheren Kontrollaktionen durchaus systematisch sind: Es werden gezielt (anhand unterschiedlicher Informationsquellen) mögliche Auslöser für konkret beschriebene, vom Fahrzeug ausgehende Gefährdungen identifiziert und die Ursachen aus der Umwelt als auslösende Umstände dokumentiert. Im Vergleich zu einem expertenbasierten Brainstorming möglicher auslösender Umstände erfolgt die Analyse also zielgerichtet und ist nicht auf Erfahrungen und Ideen einzelner Personen beschränkt. Damit stellt die STPA auch eine Möglichkeit dar, systematisch expertenbasierte Vorarbeit zu überprüfen und zu erweitern.

\section{Fazit und Ausblick} 
\label{sec:Fazit} 

In diesem Artikel beschrieben wir unsere Erkenntnisse aus der Anwendung einer STPA zur Identifikation auslösender Umstände im Kontext von SOTIF-Betrachtungen für ein automatisiertes Fahrzeug. In unserer Analyse resultierten aus einem konkreten Gefährdungsbeispiel 103 unterscheidbare Verlustszenarien, von denen wir 55 SOTIF-relevante Szenarien in kausalen Zusammenhang mit 18 beispielhaften auslösenden Umständen bringen konnten. Die Verknüpfung erfolgte über jeweils bis zu sieben Insuffizienzen, die wiederum zum Teil in mehreren Komponenten der modellierten Kontrollstruktur vorliegen können. Wir stellten entsprechend fest, dass eine deduktive Analyse nach möglichen auslösenden Umständen in der frühen Entwicklungsphase sinnvoll und mit beschriebener Strategie möglich ist. Eine ausreichende Überprüfbarkeit der Gründlichkeit einer STPA sowie deren Skalierbarkeit bleibt aber offen.

Wir haben im Zuge unserer Untersuchung festgestellt, dass der für die Beschreibung von Verlustszenarien gewählte Detaillierungsgrad große Auswirkungen auf die darauffolgende Sicherheitsanalyse und die Verknüpfung mit bereits vorhandenen Absicherungsmethoden besitzt. In unserer Beispielanwendung erfolgte die Beschreibung eher abstrakt und lässt kaum Rückschlüsse auf Szenarienelemente zu, die nicht unmittelbar Teil der untersuchten Gefährdung sind. Zukünftig wäre hier~-- insbesondere elektronisch unterstützt~-- ein größerer Detailgrad denkbar. Weitere Details in der Szenarienbeschreibung könnten einerseits die semantische Lücke zwischen Gefährdung und Ursachen beziehungsweise Auslösern besser schließen und damit grundsätzlich auch die Suche nach möglichen auslösenden Umständen unterstützen. Anderseits würde für eine systematische Untersuchung aller denkbaren Ausprägungen einer Gefährdung eine Verknüpfung mit einem gegebenenfalls bereits vorhandenen Modell der Einsatzumgebung inklusive aller Betriebsbedingungen (häufig bezeichnet als ODD~-- engl.~\foreignlanguage{english}{operational design domain}) deutliche Vorteile bieten. Zeitgleich muss festgehalten werden, dass solch eine Detaillierung von Szenarien in einer STPA grundsätzlich einer effizienten Durchführung der durch Menschen begleiteten Analyse entgegensteht. 

In unserem STPA-Beispiel wurden nicht-SOTIF-relevante Gefährdungsausprägungen analysebegleitend verworfen. Für eine Steigerung der Nutzens aus der systematischen Analyse könnte aber geprüft werden, wie eine Anbindung identifizierter Konsequenzen aus Hardware- und Softwareausfällen an Gefährdungsanalyen der Funktionalen Sicherheit (vgl. ISO~26262~\cite{international_organization_for_standardization_iso_2018}) erfolgen kann.

\begin{acknowledgement}
Wir danken den Kolleg*innen bei der \mbox{Volkswagen~AG}~-- insbesondere Jan David Schneider und Dr.-Ing. Fabian Schuldt~-- für die Unterstützung bei Durchführung und Dokumentation der hier beschriebenen Arbeiten im Zuge unseres gemeinsamen Forschungsprojekts "`Beiträge zu Sicherheit, Freigabe und Risikomanagement für ein Level-4-Fahrzeug"'. 
\end{acknowledgement}

\printbibliography[env=bibnumeric]

\addtolength{\textheight}{-7cm}

\begin{contributors}
\contributor{Robert Graubohm}{Technische Universität Braunschweig, Institut für Regelungstechnik (IfR), Hans-Sommer-Str.~66, Braunschweig, Deutschland}{graubohm@ifr.ing.tu-bs.de}{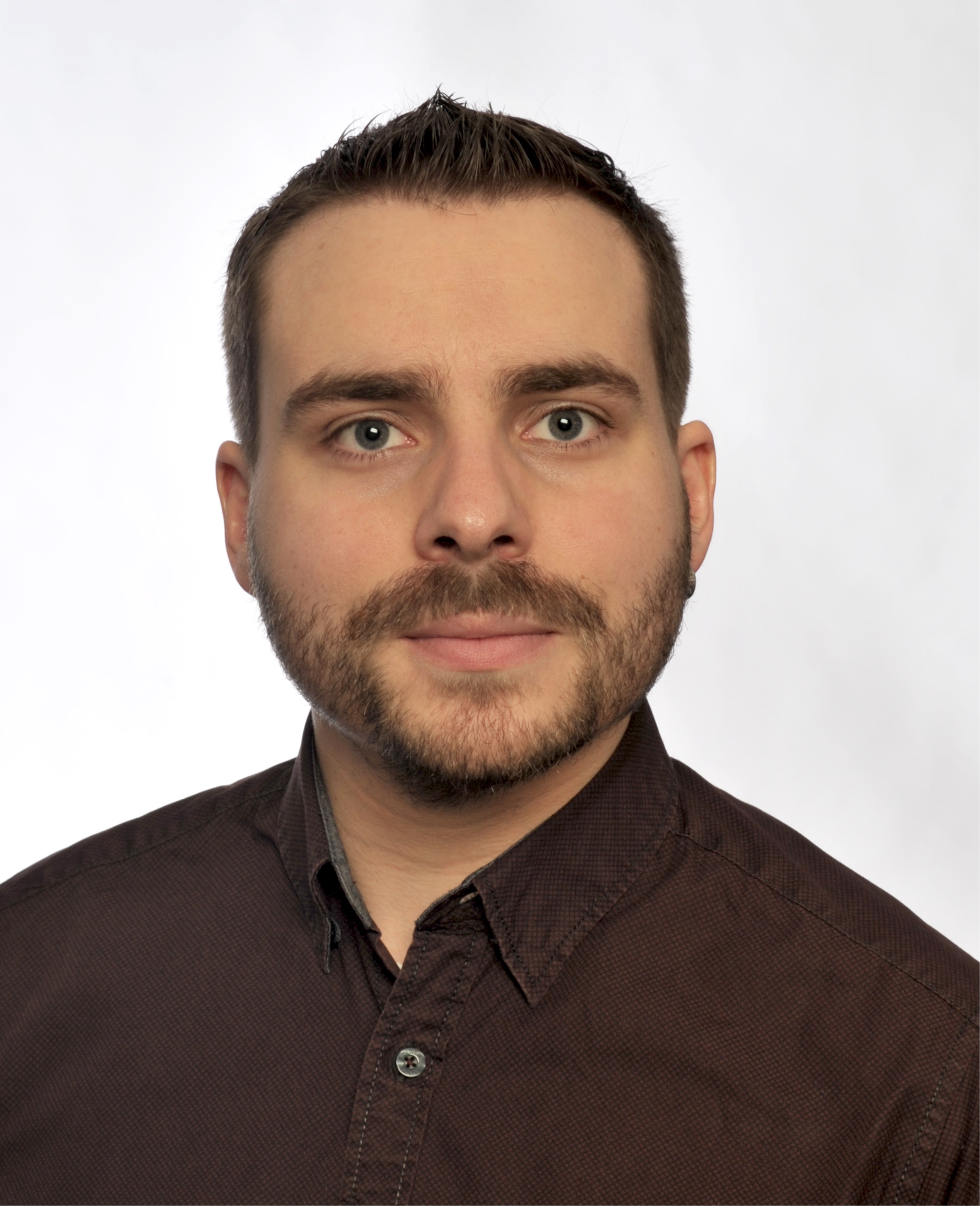}{Robert Graubohm absolvierte den M.~Sc. des Wirtschaftsingenieurwesens Maschinenbau der TU Braunschweig und MBA der University of Rhode Island. Seine Forschungsschwerpunkte sind Entwicklungsprozesse von automatisierten Fahrfunktionen und die Sicherheitskonzeption in der frühen Entwurfsphase.}
\contributor{Marvin Loba}{Technische Universität Braunschweig, Institut für Regelungstechnik (IfR), Hans-Sommer-Str.~66, Braunschweig, Deutschland}{loba@ifr.ing.tu-bs.de}{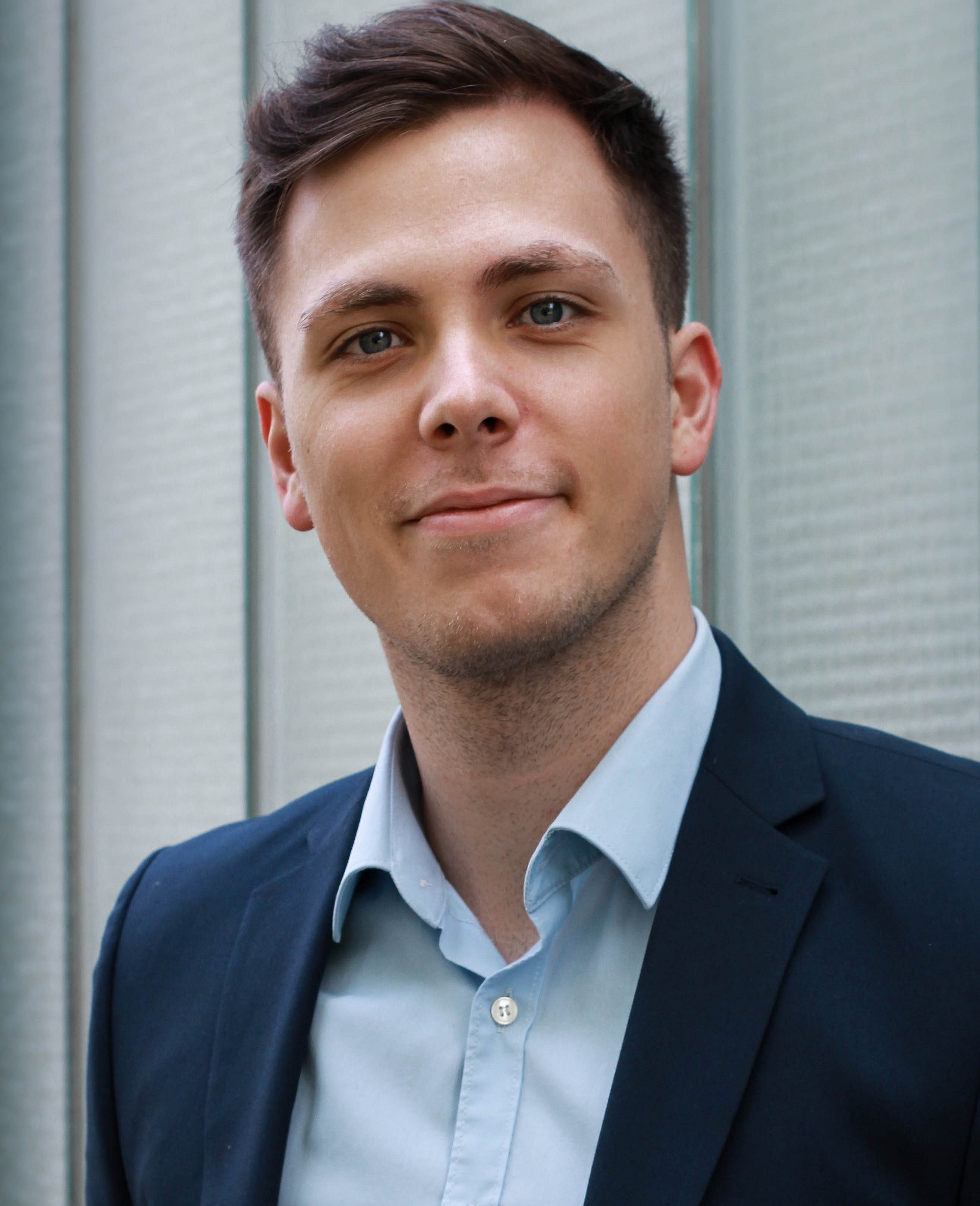}{Marvin Loba absolvierte den M.~Sc. der Elektromobilität der TU Braunschweig. Gegenstand seiner Forschung ist die Erbringung des Sicherheitsnachweises für automatisierte Straßenfahrzeuge über eine strukturierte Sicherheitsargumentation.}
\contributor{Marcus Nolte}{Technische Universität Braunschweig, Institut für Regelungstechnik (IfR), Hans-Sommer-Str.~66, Braunschweig, Deutschland}{nolte@ifr.ing.tu-bs.de}{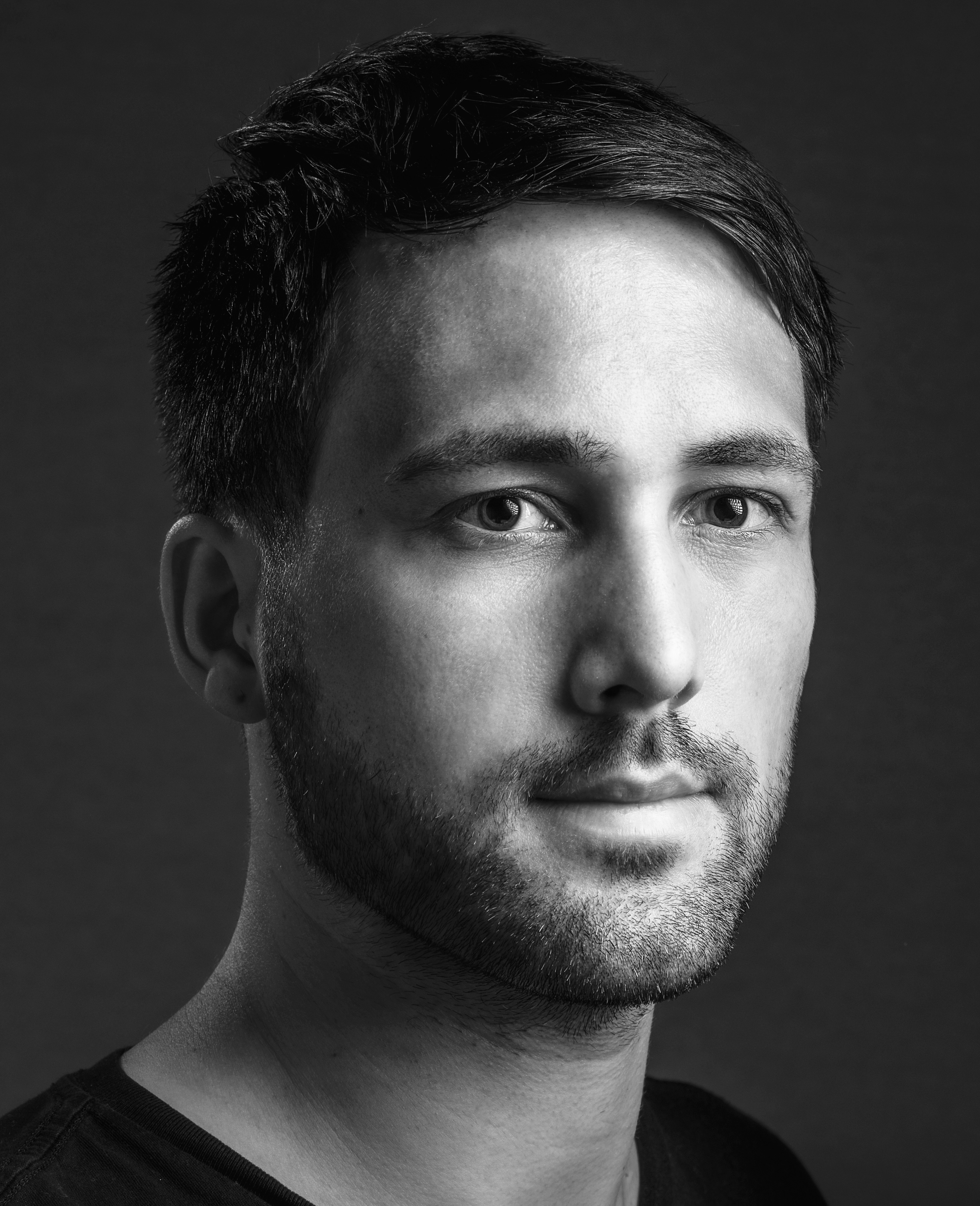}{Marcus Nolte absolvierte den M.~Sc. der Elektrotechnik an der TU Braunschweig. Sein Forschungsschwerpunkt liegt auf der Nutzung von Systems-Engineering-Ansätzen für sicheres und nachvollziehbares Verhalten automatisierter Fahrzeuge.}
\contributor{Prof. Dr.-Ing. Markus Maurer}{Technische Universität Braunschweig, Institut für Regelungstechnik (IfR), Hans-Sommer-Str.~66, Braunschweig, Deutschland}{maurer@ifr.ing.tu-bs.de}{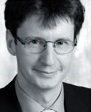}{Markus Maurer ist Professor am Institut für Regelungstechnik der TU Braunschweig. Seine Forschung konzentriert sich auf funktionale und systemische Aspekte automatisierter Straßenfahrzeuge. Von 1999 bis 2007 verantwortete er die Fahrerassistenzentwicklung bei der Audi AG.}
\end{contributors}

\end{document}